\documentclass[aps,prb,twocolumn,superscriptaddress,nofootinbib]{revtex4-1}
\pdfoutput=1
\usepackage{amssymb,graphicx,color}
\usepackage[intlimits]{amsmath}
\usepackage[english]{babel}
\usepackage[colorlinks]{hyperref}
\usepackage{comment}
\usepackage{tabularx}
\usepackage{array}
\usepackage[svgnames,table]{xcolor}
\usepackage{ragged2e}
\newcommand*{\arraycolor}[1]{\protect\leavevmode\color{#1}}
\newcolumntype{A}{>{\centering\arraybackslash \columncolor{white!50!white}}m{2.1cm}}
\newcolumntype{B}{>{\centering\arraybackslash \columncolor{white}}m{7.9cm}}
\newcolumntype{C}{>{\centering\arraybackslash \columncolor{white!50}}m{7.9cm}}
\newcolumntype{D}{>{\centering\arraybackslash \columncolor{white!42}}m}
\newcolumntype{P}[1]{>{\centering\arraybackslash}p{#1}}
\def\beq{\begin{equation}}
\def\eeq{\end{equation}}
\def\bea{\begin{eqnarray}}
\def\eea{\end{eqnarray}}
\def\barr{\begin{array}}
\def\earr{\end{array}}

\newcommand\tabitem{\makebox[1em][r]{\textbullet~}}
\begin{document}

\title{`Muhammad Ali effect' and \emph{incoherent} destruction of Wannier-Stark localization in a stochastic field}

\author{Devendra Singh Bhakuni}
\affiliation{Indian Institute of Science Education and Research Bhopal 462066 India}
\author{Sushanta Dattagupta}
\affiliation{Bose Institute, Kolkata 700054 India}
\author{Auditya Sharma}
\email{auditya@iiserb.ac.in}
\affiliation{Indian Institute of Science Education and Research Bhopal 462066 India}

\begin{abstract}
We calculate an exact expression for the probability propagator for a
noisy electric field driven tight-binding lattice.  The noise
considered is a two level jump process or a telegraph process (TP)
which jumps randomly between two values $\pm\mu$. In the absence of a
static field and in the limit of zero jump rate of the noisy field we
find that the dynamics yields Bloch oscillations with frequency $\mu$,
while with an additional static field $\epsilon$ we find oscillatory
motion with a superposition of frequencies $(\epsilon \pm \mu)$. On
the other hand, when the jump rate is ‘rapid’, and in the absence of a
static field, the stochastic field averages to zero if the two states
of the TP are equally probable ‘a-priori’. In that case we see a
delocalization effect. The intimate relationship between the rapid
relaxation case and the zero field case is a manifestation of what we
call the ‘Muhammad Ali effect’.  It is interesting to note that
even for zero static field and rapid relaxation, Bloch oscillations
ensue if there is a bias $\delta p$ in the probabilities of the two
levels. Remarkably, the Wannier-Stark localization caused by an
additional static field is destroyed if the latter is tuned to be
exactly equal and opposite to the average stochastic field $\mu\delta
p$.  This is an example of \emph{incoherent} destruction of
Wannier-Stark localization.
\end{abstract}

\maketitle 
\section{Introduction}

Anecdotes recount how when the
boxer Muhammad Ali was asked about the speed of his punches, he responded that they
were so fast that it was as if the opponent felt no punches at
all. In a variety of physical phenomena where a rapidly changing
external field is involved, this intuition of equivalence with the
corresponding zero ac field  phenomenon is indeed borne
out~\citep{bandyopadhyay2015effective,mishra2016phase,mishra2015floquet,PhysRevLett.110.200403,vcadevz2017dynamical,eckardt2005superfluid}.  We term it
the `Muhammad Ali effect'.
In the specific context of an ac electric
field applied on to a one-dimensional tight-binding lattice, the large
frequency limit can be shown to be mathematically equivalent to simply
renormalizing the hopping parameter~\cite{dunlap1986dynamic,eckardt2005superfluid}, thus
corresponding to the zero-field case.  However, for certain delicate
choices of the ratio of the amplitude and frequency, a
\emph{dynamical}
localization~\citep{PhysRevB.98.045408,dunlap1986dynamic,eckardt2009exploring,arlinghaus2011dynamic}
may be engineered via a \emph{band collapse} mechanism. On the other
hand, the zero frequency limit when the electric field is
time-independent is characterized by the familiar Bloch
oscillations~\citep{zener1934theory,krieger1986time,wannier1960wave,hartmann2004dynamics}.
Other phenomena such as coherent destruction of Wannier Stark (WS)
localization~\citep{holthaus1995random,holthaus1995ac} and super Bloch
oscillations~\citep{kudo2011theoretical,PhysRevB.86.075143,kolovsky2010dynamics,caetano2011wave}
arise when an additional static field is added on an existing
sinusoidal field. The former occurs when the static field is
resonantly tuned with the frequency of the sinusoidal field while the
latter for a slight detuning from the resonance condition.

Random disorder, in the zero electric field case is
known to localize the particle via the famous phenomenon of Anderson
localization~\cite{PhysRev.109.1492}. Since the work of Anderson, 
transport in the presence of a fluctuating environment has also been
studied~\citep{madhukar1977exact,jayannavar1982nondiffusive,girvin1979exact,kitahara1979memory,inaba1981diffusion}
both analytically and numerically. The aim here has been to understand the diffusion
of a quantum particle in the presence of dynamic disorder. This dynamic disorder originates from the
lattice vibrations where the modes of phonons are randomly excited and
the process is modeled by a stochastic
process~\citep{dattagupta2013diffusion}.
 In the presence of an electric field, disorder dephases the Bloch
oscillations depending on the strength of
disorder~\citep{drenkelforth,schulte2008dynamics,diez1998dephasing}. However
for a slowly varying disordered potential the Bloch oscillations are
known to survive~\citep{PhysRevLett.91.197402,de2005bloch}.  An increased diffusion has
been found to be the effect of scattering on the motion of a charged particle with a
time-dependent field~\citep{dunlap1988effect}.
\begin{table*}[t]
\begin{center}
\sffamily
\arrayrulewidth=1pt
\renewcommand{\arraystretch}{1.5}
\rowcolors[\hline]{3}{.!10!White}{}
\begin{tabular}{|A |B |C |}
 
    \arraycolor{black}\bfseries  &
  \arraycolor{black}\bfseries \qquad\ Deterministic $\mathbf{F = c_{0}F_{dc} + c_{1}F_{ac}}$ \newline (High frequency regime) &
  \arraycolor{black}\bfseries \qquad\ Stochastic $\mathbf{F = c_{0}F_{dc} + c_{1}F_{TP}}$ \newline (Rapid relaxation regime)\\
 DC ($c_{1}=0$) & \flushleft \hspace{0.1pt} \tabitem Bloch oscillations (WS localization). & \hspace{-71.5pt} \tabitem Bloch oscillations (WS localization).\\
 
AC/TP ($c_{0}=0$) & \hspace{-10pt}\tabitem Equivalent to no-field case (Muhammad Ali effect).\flushleft\   \tabitem  Dynamical localization with proper tuning. & \hspace{-13pt} \tabitem Equivalent to no-field case (Muhammad Ali effect). \\
 DC + AC/TP ($c_{0},c_{1}\neq 0$) & \tabitem Coherent destruction of WS localization at resonance. \flushleft \ \tabitem Super Bloch oscillations at off-resonance. &\flushleft \ \tabitem Incoherent destruction of WS localization with proper tuning of bias. \flushleft \ \tabitem Bloch oscillations with bias-dependent re-normalized frequency (another case of Muhammad Ali effect).
\end{tabular}
\caption {The table contrasts the various phenomena that arise due to ac drive and telegraph noise.}
\label{table1}
\end{center}
\end{table*}

 There are numerous experimental realizations of Bloch
  oscillations~\cite{mendez1993wannier,waschke1994experimental,
    dekorsy1994terahertz,dahan1996bloch,morsch2001bloch}. To realize
  pure Bloch oscillations, often a lot of effort is expended
  experimentally to produce clean systems since disorder and noise are
  inherent in physical systems~\cite{holthaus1995ac}. Advances in
  cold-atom technology have now made it possible to in fact control
  noise~\cite{billy2008direct,roati2008anderson,hu2015majorana} in
  order to capture special features. Therefore on the one hand, it is
  important to understand theoretically the effects of noise so that
  clever experimental techniques may be devised to get rid of them. On
  the other hand, it may be useful to understand them better so that
  they may even be exploited, given the high degree of control that
  modern technology has brought in~\cite{pichler2012noise}.   Here,
we consider the effect of a stochastic noise on top of an electric
field on the motion of the particle and focus on how the Bloch
oscillations are influenced by this type of dynamic disorder. The
particular form of the stochastic noise is the ``\emph{telegraphic
  noise}"~\citep{dattagupta2012relaxation,dattagupta2013diffusion,aharony2010retrieving,entin2017heat},
where the noise consists of jumps randomly between two levels
$\pm\mu$.  Telegraph noise is one of the simplest realizations of
fluctuations in the battery. When such a telegraphic noise term
appears as fluctuations in the site energies without any linear
variation (the limit when the electric field is zero), exact
analytical results for the diffusion coefficient have been
obtained~\citep{inaba1981diffusion}. Also the effect of
  noise on dynamical localization has been
  studied~\cite{suqing2000effect}. Here, we consider the case where
the noise term acts as fluctuations to an electric field. This noise
can also be thought of as an aperiodic form of a square wave driving
(periodic square wave driving with proper tuning can yield dynamic
localization
~\citep{dunlap1988dynamic,luitz2017absence}).  Perfect
  periodic drive is impossible to
  achieve~\cite{rieder2018localization,sieberer2018statistical} in
  realistic experimental situations and therefore it is important to
  study the effects of noisy
  drive~\citep{vcadevz2017dynamical,hatami2016quasiperiodic,cai2017universal,nandy2017aperiodically}.

  The central findings (Table~\ref{table1}) of our Letter are as follows. For a stochastic
  electric field characterized by telegraph noise, we find the exact
  expression for the probability propagator $\mathcal{P}_m(t)$,
  defined as the probability of a particle to remain at site $m$ at
  any time $t$ given that it was at the origin at $t=0$. The limit of
  the rapidly changing stochastic field is given particular
  emphasis. Denoting the bias in the probabilities of the two levels
  of the field to be $\delta p$, we show that this is equivalent to an
  effective dc field of $\mu\delta p$, yielding Bloch oscillations
  with frequency $\mu\delta p$ (although these oscillations are
  exponentially damped in the infinite time limit). If an additional
  static field is present, we recover Bloch oscillations with a
  renormalized frequency in the rapid relaxation limit - this is another 
  instance of the Muhammad Ali effect. Remarkably, by choosing
  the additional static field to have a precise magnitude, a
  destruction of WS localization~\cite{holthaus1995random,holthaus1995ac} may be
  engineered. Since no frequency is involved in the present context,
  and rather the noise may be a result of connection to a bath, this
  may be termed an \emph{incoherent} destruction of WS
  localization. When the two levels of the stochastic field are
  equiprobable ($\delta p =0$), we recover the well-known scenario
  that the rapid relaxation limit is equivalent to the zero-field
  limit: the `Muhammad Ali effect'. A complementary numerical approach is used to independently
  verify our findings.

\section{Model Hamiltonian and probability propagator}

The Hamiltonian for a $1\text{D}$ tight binding model with a time dependent electric field is
\begin{equation}
H=-\frac{\Delta}{4}\sum_{n=-\infty}^{\infty}c_{n}^{\dagger}c_{n+1}+c_{n+1}^{\dagger}c_{n}+\mathcal{F}(t)\sum_{n=-\infty}^{\infty} n c_{n}^{\dagger}c_{n},
\end{equation}
where $\mathcal{F}(t)$ is the electric field. The lattice constant is kept at unity and natural units
($\hbar=e=1$) are adopted for all the calculations.  For a
constant electric field, the dynamics gives the
well known Bloch oscillations, while a periodic driving can give rise
to dynamical localization when the amplitude and frequency are tuned
appropriately. Here, we consider the case where the time dependent
electric field is described by a two state jump process or a
\emph{telegraph process}.

It is useful to define the unitary operators $\hat{K},\hat{K}^{\dagger}$ and $\hat{N}$~\citep{hartmann2004dynamics}, and their operations on the state $|n\rangle$ as 
\begin{eqnarray}
\hat{K}=\exp{\left(-i\hat{\kappa}\right)}=\sum_{n=-\infty}^{\infty}|n\rangle\langle n+1|, \ \hat{K}|n\rangle=|n-1\rangle \nonumber\\ 
\hat{K^{\dagger}}=\exp{\left(i\hat{\kappa}\right)}=\sum_{n=-\infty}^{\infty}|n+1\rangle\langle n|, \ \hat{K}^\dagger|n\rangle=|n+1\rangle\nonumber\\
\hat{N}=\sum_{-\infty}^{\infty}n|n\rangle\langle n|.\qquad\qquad\qquad\qquad\qquad
\end{eqnarray}
These operators follow the commutation rules
\begin{equation}\label{commutation}
\left[\hat{K},\hat{N}\right]=\hat{K},\ \left[\hat{K}^\dagger,\hat{N}\right]=-\hat{K}^\dagger,\ \left[\hat{K},\hat{K}^\dagger\right]=0.
\end{equation} 
The eigenvectors of $\hat{K}$ are the Bloch states $|\kappa\rangle$
with eigenvalues $e^{i\kappa}$. The connection between the Wannier
basis and the Bloch basis is given by $|k\rangle =
\sqrt{\frac{1}{2\pi}}\sum_n e^{-ink}\ |n\rangle$ and $|n\rangle =
\frac{1}{2\pi }\int_{-\pi}^{\pi} \ dk \ e^{ink}\ |k\rangle $.

In terms of these new operators, the tight-binding Hamiltonian can be written as 
\begin{equation}
\textit{$\hat{H}$}(t)=V^+ + H_0(t),
\end{equation}
where $V^\pm=-\frac{\Delta}{4}\left(\hat{K}\pm\hat{K}^\dagger\right)$ and $H_0(t)=\mathcal{F}(t)\hat{N}$.

The time evolution of the density matrix $\rho$ in Heisenberg picture is given by 
\begin{equation}
\frac{\partial\rho}{\partial t}= -i\left[H(t),\rho(t)\right].
\end{equation}
By considering the transformation $\tilde{\rho}(t)=e^{i\int_{0}^{t}H_{0}(t^\prime) dt^\prime}\rho(t)e^{-i\int_{0}^{t}H_{0}(t^\prime) dt^\prime}$, the equation of motion for $\tilde{\rho}(t)$ can be written as
\begin{equation}\label{eq10}
\frac{\partial\tilde{\rho}}{\partial t}= -i\left[\tilde{V}^{+}(t),\tilde{\rho}(t)\right],
\end{equation}
where $\tilde{V}^{+}(t)=e^{i\int_{0}^{t}H_{0}(t^\prime) dt^\prime}V^{+}e^{-i\int_{0}^{t}H_{0}(t^\prime) dt^\prime}$. The time evolution of $\tilde{\rho}$ can now be solved to
\begin{equation}
\tilde{\rho}(t)=e^{-i\int_{0}^{t}\tilde{V}^{+}(t^\prime) dt^\prime}\rho(0)e^{i\int_{0}^{t}\tilde{V}^{+}(t^\prime) dt^\prime},
\end{equation}
where $\tilde{\rho}(0)=\rho(0)=|0\rangle\langle 0|$. It turns out that $[\tilde{V}^{+}(t),\tilde{V}^{+}(t^{'})]$ even for $t\neq t^{'}$, and therefore no complicated time-ordering is essential.

Using the Baker -Campbell-Hausdorff (BCH) formula $e^XYe^{-X}=Y+\left[X,Y\right]+\frac{1}{2!}\left[X,\left[X,Y\right]\right]+....$, and the commutation relations (Eqn.~\ref{commutation}), we can simplify the effective Hamiltonian which governs the dynamics of the density matrix $\tilde{\rho}(t)$ as
\begin{equation}
\tilde{V}^{+}(t)=V^{+}\cos(\eta(t))+iV^{-}\sin(\eta(t)),
\end{equation} 
where $\eta(t)=\int_{0}^{t} \mathcal{F}(t^\prime) \ dt^\prime$. Substituting the expressions of $V^{+}$ and $V^{-}$, we get
\begin{equation}
\tilde{V}^{+}(t)=-\frac{\Delta}{4}\left(\hat{K}^\dagger e^{i\eta(t)}+\hat{K} e^{-i\eta(t)}\right).
\end{equation}
It can be seen that the effective Hamiltonian has the time dependence
appearing as a phase term and it can be easily diagonalized in the
momentum basis. In the $k$ representation, $\tilde{V}^{+}(t)$ can be
expressed as
\begin{equation}
\langle k|\tilde{V}^{+}(t)|k^{\prime}\rangle =-\frac{\Delta}{4}\delta(k-k^\prime)\left[e^{ik+i\eta(t)}+e^{-ik-i\eta(t)}\right].
\end{equation}
Furthermore, the transformed density matrix $\tilde{\rho}(t)$ can also
be written in $k$ basis as
\begin{equation}\label{rho}
\langle k|\tilde{\rho}(t)|k^{\prime}\rangle = e^{-i\int_{0}^{t}dt^\prime V_{k}^{+}(t^\prime)}\langle k|0\rangle\langle 0|k^\prime\rangle e^{i\int_{0}^{t}dt^\prime V_{k^\prime}^{+}(t^\prime)}, 
\end{equation}
where $V_{k}^{+}(t) =-\frac{\Delta}{4}\left[e^{i(k+\eta(t))}+e^{-i(k+\eta(t))}\right]$.

In Wannier space the probability propagator is given by
\begin{equation}
\label{eqn:prop}
\mathcal{P}_m(t)=\langle m |\rho(t)|m\rangle = \langle m |\tilde{\rho}(t)|m\rangle,
\end{equation}
where we have used the fact that $H_0$ is diagonal in Wannier basis. Going into the momentum basis the expression for the probability can be simplified to
\begin{eqnarray}\label{eq13}
\mathcal{P}_m(t)=\int\int dk\ dk^\prime\  \langle m|k\rangle\langle k|\tilde{\rho}(t)|k^\prime\rangle\langle k^\prime|m\rangle,
\end{eqnarray}
which using Eq.~\ref{rho} takes the simplified form as :
\begin{eqnarray}\label{eq13}
\mathcal{P}_m(t)=\left(\frac{1}{2\pi}\right)^2\int_{-\pi}^{\pi} dk \int_{-\pi}^{\pi} dk^{\prime} e^{-i(k-k^{\prime})m}\qquad\qquad\qquad \nonumber \\ \times \ e^{-i\int_{0}^{t} dt^{\prime}(V_{k}^{+}(t^{\prime})-V_{k^{\prime}}^{+}(t^{\prime}))},\qquad\qquad
\end{eqnarray}
The mean squared width of the wave-packet is then expressed in terms of the probability propagator as
$\sigma^2(t)=\langle m^2\rangle = \sum_{m} m^2\mathcal{P}_m(t)$.

\section{Effect of random telegraph noise}

 The particular form of the field is taken as a telegraph noise where
 electric field is time-dependent and randomly fluctuates between two
 levels $\pm\mu$. Let $\sigma$ and $\tau$ be the rate of
switching from level $+\mu$ to $-\mu$ and $-\mu$ to $+\mu$ respectively. The
probability of being at any time in state $+\mu$ is given by
$p_{+} = \tau/(\tau+\sigma)$, whereas the probability of being in state $-\mu$ is
$p_{-} = \sigma/(\tau+\sigma)$. It is useful to define $\lambda = \sigma+\tau$.

 A clever way to make progress is to elevate  $i\eta(t)=i\int_{0}^{t}
 \mathcal{F}(t^\prime) \ dt^\prime$ to a $2\times 2$ matrix~\citep{blume1968stochastic} 
\begin{equation}\label{eq21}
i\eta(t)=it\epsilon.\mathcal{I}+it\mu\sigma_z+\lambda t W,
\end{equation}
in which $\mathcal{I}$ is the identity matrix, $\sigma_z$ is the Pauli $z$ matrix, and the relaxation matrix~\cite{dattagupta2012relaxation,aharony2010retrieving} is defined as
\begin{equation}
  W =  \begin{bmatrix}
    -p_- & p_+\\
    p_- & -p_+ \\
  \end{bmatrix}
  =\lambda
  \begin{bmatrix}  
   -\frac{\sigma}{\tau+\sigma} & \frac{\tau}{\tau+\sigma}\\
    \frac{\sigma}{\tau+\sigma} & -\frac{\tau}{\tau+\sigma} \\
  \end{bmatrix}.\\
\end{equation}
As a consequence of this operation, the probability propagator (Eqn.~\ref{eqn:prop}) is also now a $2\times 2$ matrix.
The first term added in Eqn.~\ref{eq21} is to account for the static electric field
$\epsilon$ and the two stochastic states are $|+\rangle = \begin{pmatrix}1\\0\end{pmatrix},|-\rangle = \begin{pmatrix}0\\1\end{pmatrix}$
corresponding to the fields $+\mu$ and $-\mu$ respectively. Eqn.~\ref{eq21} can be decomposed in terms of Pauli matrices as
\begin{equation}\label{eq22}
i\eta(t)= -t(\gamma-i\epsilon).\mathcal{I} + t\sigma_z(\gamma\delta p + i\mu) + \gamma t(\sigma_x +i\delta p\sigma_y),
\end{equation}
where $\gamma=\frac{\lambda}{2}$ and $\delta p=(p_+-p_-)$.
The exponential of Eqn.~\ref{eq22} can be written in a compact form: $
e^{i\eta(t)}= e^{-t(\gamma-i\epsilon)} \ e^{t(\textbf{h}.\boldsymbol{\sigma})},$
where $h_x=\gamma, h_{y}=i\gamma\delta p$ and  $h_z=(\gamma\delta p + i\mu)$ and $|\textbf{h}|=\sqrt{\gamma^2 - \mu^2 + 2i\gamma\mu\delta p}=\nu$. 
Using the Pauli spin identity: $e^{i(\textbf{a}.\boldsymbol{\sigma})}=\mathcal{I}\cos{|\textbf{a}|} + i(\hat{n}.\vec{\sigma})\sin{|\textbf{a}|}$,
the above expression can be written as
\begin{eqnarray}\label{eq24}
e^{i\eta(t)}=\frac{1}{2} e^{-t(\gamma-i\epsilon)}\left[e^{\nu t}(1+\hat{h}.\vec{\sigma}) + e^{-\nu t}(1+\hat{h}.\vec{\sigma})\right].\quad\  
\end{eqnarray}
Similarly, an equation for the complex conjugation can be written with $h_{x}^{\prime}=\gamma, h_{y}^\prime=-i\gamma\delta p, h_{z}^{\prime}=\gamma\delta p-i\mu$. 

After some lenghty calculations (detailed in the Appendix), the exponential part of Eqn.~\ref{eq13} can be written as
\begin{eqnarray}
e^{-i\int_{0}^{t} dt^{\prime}(V_{k}^{+}(t^{\prime})-V_{k^{\prime}}^{+}(t^{\prime}))}=i[g_0(t)\mathcal{I} + \alpha(t)\sigma_x+ i\delta p \alpha(t)\sigma_y \nonumber\\ +\  (\delta p \alpha(t) + \beta(t))\sigma_z],\qquad\quad\quad
\end{eqnarray}
where the complicated expressions for $g_{0}(t),\alpha(t)$ and $\beta(t)$ are relegated to the Appendix.
Finally, we have the compact form 
\begin{equation}
e^{-i\int_{0}^{t} dt^{\prime}(V_{k}^{+}(t^{\prime})-V_{k^{\prime}}^{+}(t^{\prime}))}=e^{ig_0(t)} \ e^{i(\textbf{H}.\boldsymbol{\sigma})},
\end{equation}
where $H_x=\alpha(t),\ H_y=i\delta p \alpha(t)$ and $H_z=\delta p \alpha(t) + \beta(t)$ and $|\textbf{H}| = \sqrt{\alpha^2(t) + \beta^2(t)+2\delta p \alpha(t)\beta(t)}$.
Again using the Pauli spin identity, we get
\begin{equation}
e^{i(\textbf{H}.\boldsymbol{\sigma})}=\left[\mathcal{I}.\cos{|\textbf{H}|} + i(\hat{H}.\vec{\sigma})\sin{|\textbf{H}|}\right].
\end{equation}
For the final expression of probability, we
need to calculate the restricted average
$\bar{\mathcal{P}}_{m}(t)=\sum_{ab}p_a(b|\mathcal{P}_m(t)|a)$~\cite{blume1968stochastic}. The
averages of $\sigma_{y}$ and $\sigma_{z}$ give $i\delta p$ and $\delta
p$ respectively, whereas $\sigma_x$ averages to unity.  While
$\mathcal{P}_m(t)$ is a matrix, the average $\bar{\mathcal{P}}_m(t)$
is just a number. The final expression for the average probability can
be written as
\begin{eqnarray}\label{prob1}
\bar{\mathcal{P}}_m(t) = \left(\frac{1}{2\pi}\right)^2\int_{-\pi}^{\pi} dk \int_{-\pi}^{\pi} dk^{\prime} \  e^{-i(k-k^{\prime})m}\ e^{ig_0(t)}\times\qquad\quad\nonumber \\ \left [\cos{|\textbf{H}|} + i\sin{|\textbf{H}|}\frac{\alpha(t)}{|\textbf{H}|} +i\delta p\sin{|\textbf{H}|}\frac{\beta(t)}{|\textbf{H}|} \right],\qquad
\end{eqnarray}
which is one of our key results. It is straightforward to verify the
well-known results for the zero field ($\epsilon=0$) and static field
($\epsilon\ne 0$) case where $\mu$ vanishes, hence $\beta(t)=0$ and
$\nu\rightarrow \gamma$. In the former case the probability propagator
decays in time and the mean squared width becomes unbounded in
time. Hence, an initially localized particle will delocalize. In the
latter case of static field, both the probability and the mean squared
width are bounded and exhibit the familiar Bloch oscillations with
frequency $\omega_{B} = \epsilon$.  Further, considering the effect of
telegraph noise the zero relaxation limit where $\gamma=0, \nu=i\mu$
(and hence $\alpha(t)=0$) is straightforward. A simplification of the
probability propagator in this limit yields a superposition of
probabilities for the two `static fields' $(\epsilon\pm\mu)$.

The rapid relaxation condition $\gamma >> \mu,\epsilon$ is the core
emphasis of our Letter, and will be imposed in the rest of the discussion
ahead.  In this limit, $\alpha^2(t)>>\beta^2(t)$ and an expansion of
$|\textbf{H}|$, $\frac{\alpha(t)}{|\textbf{H}|}$ and
$\frac{\beta(t)}{|\textbf{H}|}$ simplifies the integrand of the
probability propagator (Eqn.~\ref{prob1}) as
\begin{eqnarray}\label{eq40}
 e^{ig_0(t)}\left [\cos{|\textbf{H}|} + i\sin{|\textbf{H}|}\frac{\alpha(t)}{|\textbf{H}|} +i\delta p\sin{|\textbf{H}|}\frac{\beta(t)}{|\textbf{H}|} \right]\qquad\qquad\nonumber\\
\approx e^{ig_0(t)+i\alpha(t)+i\delta p\beta(t)} e^{i\left(\frac{\beta^2(t)}{2\alpha(t)}\right)}.\qquad
\end{eqnarray}
We consider separately the cases where both the levels of the
stochastic field are equally probable ($\delta p =0$) and where one
level is more probable than the other $(\delta p \ne 0)$.

With $\delta p=0$, and $\gamma>>\mu$, the expression for $\nu$ can be
expanded upto $\mathcal{O}(\frac{\mu^2}{\gamma})$ as
$\nu=\sqrt{\gamma^2-\mu^2}\approx\gamma-\frac{\mu^2}{2\gamma}$.
\begin{figure}[t]
\includegraphics[scale=0.9]{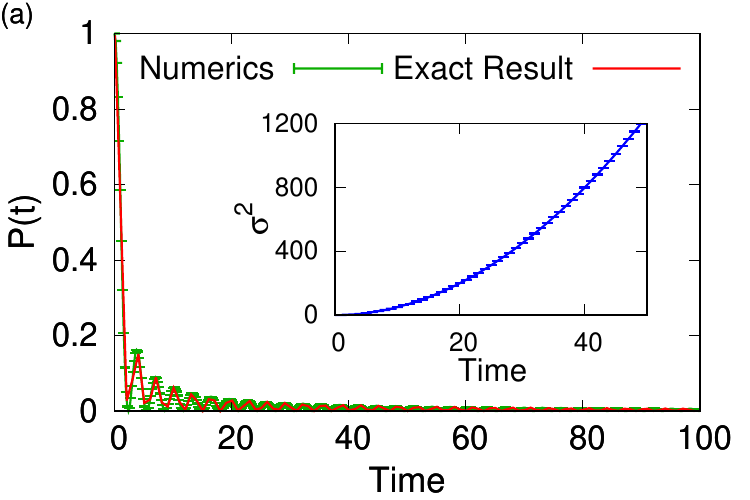}
\includegraphics[scale=0.9]{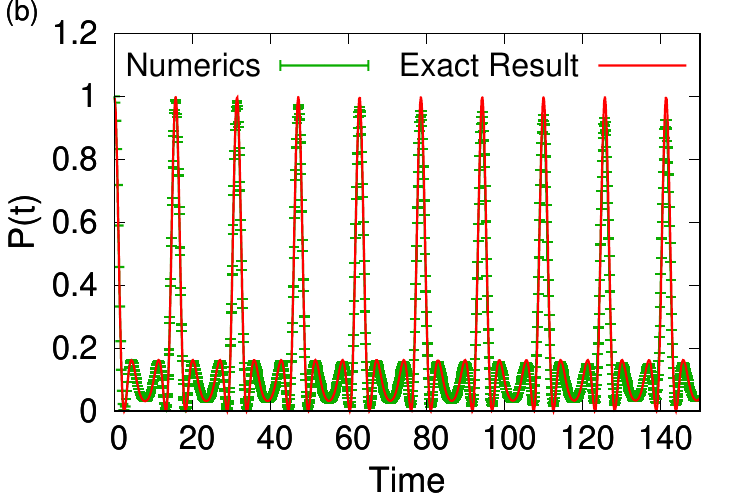}
\caption{The return probability of an initially localized wave-packet
  $(\delta_{m,0})$ from the exact calculation and exact numerics. Here 
  we present data for the case of zero bias ($\delta p=0$) in the rapid relaxation regime
  ($\sigma=\tau=100$) with $\Delta=2.0$. (a) Muhammad Ali effect in the zero static field limit $\epsilon=0.0$. The inset shows the unbounded
  growth of mean squared width, analogous to the zero-field scenario. 
  (b) Bloch oscillations in the finite
  static field limit $\epsilon=0.4$. In both the figures the numerics are performed
  for a system of size $L=400$ with averaging carried out over $100$
  realizations of the disorder.}
\label{fig3}
\end{figure}
For the zero static field case ($\epsilon = 0$), the expressions for $g_0(t),\alpha(t)$ and $\beta(t)$ can be written as (upto $\mathcal{O}(\mu/\gamma)$)
\begin{eqnarray}
g_0(t)\approx\eta_+\left(\cos k-\cos k^\prime\right),\ 
\alpha(t)\approx\eta_-\left(\cos k-\cos k^\prime\right)\nonumber\\
\beta(t)\approx-\frac{\mu}{\gamma}\eta_-\left(\sin k-\sin k^\prime\right),\qquad\qquad\quad
\end{eqnarray}
where $\eta_\pm(t)=\frac{\Delta}{4}\frac{1}{2\gamma}\left[2\gamma t\pm(1-e^{-2\gamma t})\right]$.
Substituting the values of $g_0(t),\alpha(t)$ and $\beta(t)$ and taking the long time limit, we get
\begin{eqnarray}
e^{ig_0(t)+i\alpha(t)} e^{i\frac{\beta^2(t)}{2\alpha(t)}}\approx e^{i\frac{\Delta_{\text{eff}}t}{2}(\cos k-\cos k^\prime)},
\end{eqnarray}
where
$\Delta_{\text{eff}}=\Delta\left[1+\frac{1}{8}\left(\frac{\mu}{\gamma}\right)^2
  \left(\frac{\sin k-\sin k^\prime}{\cos k-\cos k^\prime}\right)^2\right]$.
Hence in this limit, the effect is identical to the case of no field, a phenomenon we have called `Muhammad Ali effect'. This is the case
where the electric field is so rapidly fluctuating between $\pm \mu$,
that for all practical purposes the system feels no effect at
all. This effect is shown in Fig.~\ref{fig3}, where the probability
propagator and the mean squared width of the wave-packet are plotted
with time. The return probability decays in time and the
wave-packet width becomes unbounded signifying the delocalization
of an initially localized wave-packet. In the presence of the static field ($\epsilon \neq 0$), we have the approximation
\begin{eqnarray}\label{eq43}
g_0(t)+\alpha(t) \approx \frac{\Delta}{2\epsilon}\left[e^{-\frac{\mu^2}{2\gamma}t}\sin(k+\epsilon t)- \sin k \right]\quad\nonumber\\-\frac{\Delta}{2\epsilon}\left[e^{-\frac{\mu^2}{2\gamma}t}\sin(k^\prime+\epsilon t)- \sin k^\prime \right].
\end{eqnarray}
In the limit $\gamma>>\mu$, the ratio $\frac{g_{3}^2(t)}{2g_2(t)}$
becomes very small and can be neglected. Also the term
$e^{-\frac{\mu^2}{2\gamma}t}$ becomes unity, unless $t$ is very large. So in this limit, one
obtains Bloch oscillations with frequency $\epsilon$ for small times (Fig.~\ref{fig3}); however the rapidly fluctuating noise causes in
the long time limit for these oscillations to damp out exponentially.

\begin{figure}[t]
\includegraphics[scale=0.9]{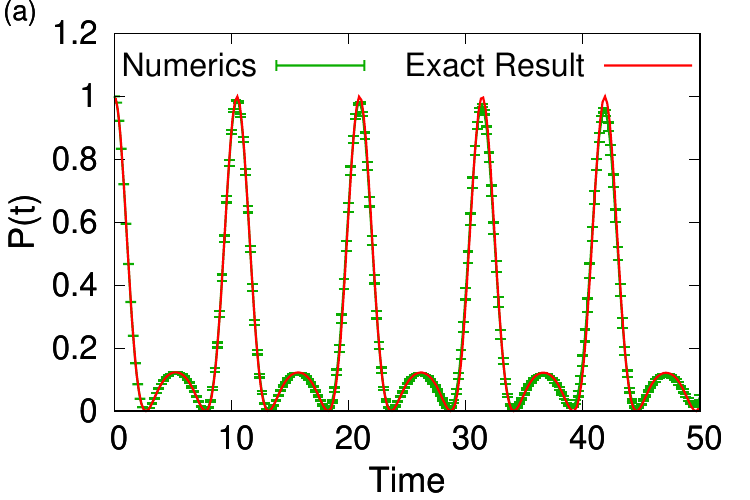}
\includegraphics[scale=0.9]{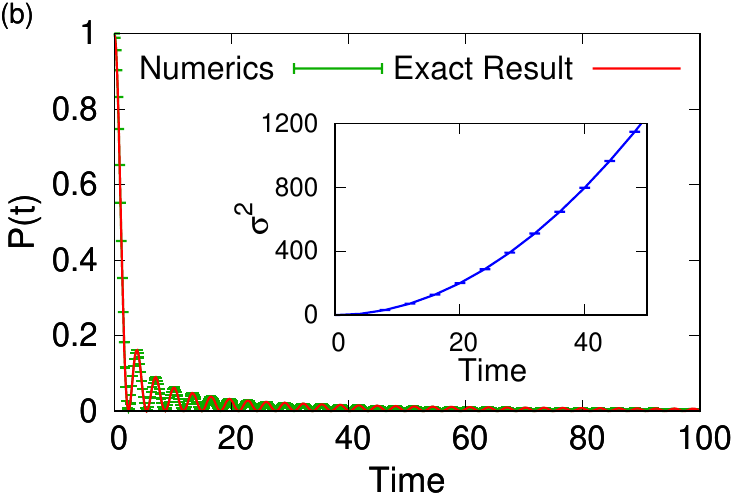}
\caption{The return probability of a wave-packet initially localized
  state at the center of the chain ($m=0$), from exact calculation and
  exact numerics. (a) Bloch oscillations with renormalized frequency
  ($\epsilon+\mu\delta p$) are seen in the rapid relaxation regime
  ($\sigma=50,\ \tau=150$) and with bias $\delta p=0.5$ and static field
  $\epsilon=0.5$. (b) Incoherent destruction of WS
  localization by the stochastic field for
  $\epsilon=0.1,\ \sigma=150,\ \tau=50,\ \delta p=-0.5$. The inset
  shows the unbounded growth of mean squared width. In both the figures, the other
  parameters are: $\mu=0.2,\ L=400,\ \Delta=2.0,\ dt=0.01,$ and the exact numerics
  average over $100$ realizations of disorder. }
\label{fig5}
\end{figure}

Another interesting case of rapid relaxation arises when the two
levels are not equiprobable ($\delta p\neq 0$). Here
$\nu=\sqrt{\gamma^2 - \mu^2 + 2i\gamma\mu\delta p}$. We can expand $\nu$ upto 
$\mathcal{O}\left(\frac{\mu^2}{\gamma^2}\right)$ as
$\nu=\gamma\left(1+i(\frac{\mu}{\gamma})\delta p-\frac{\mu^2}{2\gamma^2}\right)$ and 
$\gamma -\nu = \frac{\mu^2}{2\gamma^2} - i\mu\delta p$. With these approximations and defining $\xi = \epsilon +\mu\delta p$, the exponent of the first part of Eqn.~\ref{eq40} can be simplified to
\begin{eqnarray}
g_0(t)+\alpha(t)+\delta p \beta(t) \approx \frac{\Delta}{2\xi}\left[e^{-\frac{\mu^2}{2\gamma}t}\sin(k+\xi t)- \sin k \right]\nonumber\\-\frac{\Delta}{2\xi}\left[e^{-\frac{\mu^2}{2\gamma}t}\sin(k^\prime+\xi t)- \sin k^\prime \right].\qquad
\end{eqnarray}
The above expression is similar to Eqn.~\ref{eq43} with $\epsilon$
replaced by $\xi$. Hence, Bloch oscillations with the average field
and frequency $\xi$ appear, which in the long time limit damp out
exponentially. Also, unlike the case of $\delta p=0$, Bloch
oscillations with frequency $\mu\delta p$ arise even in the zero
static field case. Tuning the bias: $\delta p = -\frac{\epsilon}{\mu}$
in order to precisely cancel the effect of the static field, causes
the average electric field to become zero, as a consequence of which
Bloch oscillations are destroyed. This can be termed as
\emph{incoherent} destruction of WS localization as no frequency is
involved in this scenario. This is to be contrasted with
\emph{coherent} destruction of WS
localization~\citep{holthaus1995random,holthaus1995ac}, where a
resonant tuning of the drive provides the mechanism in a system that
is subjected to a combined dc and time periodic ac field. The incoherent destruction of localization  here is to be seen as a contrast with the `Muhammad Ali effect'. All these effects are
plotted in Fig.~\ref{fig5}, where the return probability and the mean
squared width of the wave-packet are given as a function of time. The
details of the numerical generation of the telegraph noise are given
in the next section.

\section{Numerical implementation of telegraph noise }
The different cases considered above for the telegraphic noise can be
verified independently from an exact numerical approach. The numerical
approach involves the implementation of telegraph noise followed by
the diagonalization of the Hamiltonian at each instant of time. The
probability propagator can then be calculated by looking at the
dynamics of an initial state.

For the numerical generation of the telegraph noise we follow
Refs.~\onlinecite{machlup1954noise,barik2006langevin,xu2012stochastic,dattagupta2012relaxation,dattagupta2013diffusion}.
The method works as follows: Let $\sigma$ and $\tau$ be the rate of
switching from level $a$ to $b$ and $b$ to $a$ respectively. The
probability of being at any time in state $a$ is given by
$\tau/(\tau+\sigma)$, whereas the probability of being in state $b$ is
$\sigma/(\tau+\sigma)$. Furthermore, let $w_{ij}=(i|W|j)$ with
$i,j=\{a,b\}$ be the matrix elements of the relaxation matrix which
gives the transition rate to jump from a state $j$ to $i$.  The
condition of detailed balance implies
\begin{equation}\label{detailbal}
p_{b} (a|W|b) = p_{a} (b|W|a),
\end{equation}
where $p_a$ and $p_b$ are the probability to remain in state $a$ and $b$ respectively. 
Invoking conservation of probability along with Eqn.~\ref{detailbal}, the matrix element of the relaxation matrix can be expressed as
\begin{equation}
w_{ab}=\lambda p_a,  \ \ w_{ba}=\lambda p_{b}, 
\end{equation}
where $\lambda=w_{ab}+w_{ba}$. 

The relaxation matrix can thus be written as
\begin{equation}
W=\lambda
  \begin{bmatrix}
  
   -p_{b} & p_{a}\\
    -p_{b} & -p_{a} \\
  \end{bmatrix}\\.
\end{equation}
By substituting the values of $p_a$ and $p_b$, the relaxation matrix $W$ can be expressed as 
\begin{equation}
W=\lambda
  \begin{bmatrix}
  
   -\frac{\sigma}{\tau+\sigma} & \frac{\tau}{\tau+\sigma}\\
    \frac{\sigma}{\tau+\sigma} & -\frac{\tau}{\tau+\sigma} \\
  \end{bmatrix}\\,
\end{equation}
where, $\lambda=\tau+\sigma$. The difference of the probabilities between the two levels can be extracted as: $\delta p=\frac{\tau-\sigma}{\tau+\sigma}$.

Also the various conditional probabilities can be expressed in terms of the elements of the relaxation matrix as follows~\citep{barik2006langevin,xu2012stochastic}:
\begin{eqnarray}
P_{aa}=P\left(a,t_{n+1}|a,t_n\right) = \frac{\sigma}{\tau + \sigma} + \frac{\tau}{\tau + \sigma}\exp\left(-(\tau + \sigma)dt\right)\nonumber\\
P_{ba}=P\left(a,t_{n+1}|b,t_n\right) = \frac{\sigma}{\tau + \sigma} - \frac{\sigma}{\tau + \sigma}\exp\left(-(\tau + \sigma)dt\right)\nonumber\\
P_{bb}=P\left(b,t_{n+1}|b,t_n\right) = \frac{\tau}{\tau + \sigma} + \frac{\sigma}{\tau + \sigma}\exp\left(-(\tau + \sigma)dt\right)\nonumber\\
P_{ab}=P\left(b,t_{n+1}|a,t_n\right) = \frac{\tau}{\tau + \sigma} - \frac{\tau}{\tau + \sigma}\exp\left(-(\tau + \sigma)dt\right)\nonumber\\
\end{eqnarray}

Finally, the numerical simulation is done as follows. Let the starting
state be $a$. A random number between $0$ and $1$ is generated from the computer, 
and is compared against the conditional probability $P_{aa}$. If the conditional 
probability is greater than the random number, the next state will remains $a$, otherwise the next state
will be changed to $b$. If the state changes to $b$, then for the next
time, a random number is again generated and contrasted against the
conditional probability $P_{ba}$. If this conditional probability is
greater than the random number, the next state is taken as $a$ else it
will remain $b$. If the starting state is $b$, the random number is compared against the
conditional probability $P_{bb}$. Again if this conditional
probability is greater than the random number, the next state will
remain $b$, otherwise it will be changed to $a$. If a flip happens to
$a$, then a random number is generated and compared against the
conditional probability $P_{ab}$. If this conditional probability is
greater than the random number, the next state will flip to $b$, else
it will remain $a$. This process is repeated in time units of length $dt$ until the final
time is reached. The different cases of the telegraphic noise can then
be generated by setting the values $\sigma$ and $\tau$.

\section{Summary and Conclusions }

To summarize, we studied the effect of an electric field subjected to
random telegraphic noise on a nearest-neighbor tight-binding
chain. Our first result is the derivation of an exact general
expression for the probability propagator, which is then employed to
illuminate several special cases.  As expected, in the zero relaxation
case, the probability shows oscillatory behavior, with a superposition
of the frequencies `$\epsilon\pm\mu$'. The rapid relaxation scenario
forms the core emphasis of our work, and may be subdivided into two
cases: one where the rates for the two levels are the same and the
other where one level has greater lifetime than the other. In the
former case, a delocalization effect is obtained in zero static field
and Bloch oscillations in the presence of a static field. We identify this limit as a manifestation of what we call the `Muhammad Ali effect'. In the latter case , a finite difference in the probabilities of the two levels renormalizes the Bloch frequency to
$\omega_{B}=\epsilon + \mu\delta p$. A precise tuning of the bias
$\delta p$ leads to \emph{incoherent} destruction of WS
localization. The exact results are also verified by an independent
numerical approach as well.

\section*{Acknowledgements}
A.S is grateful to Arul Lakshminarayan for narrating the Muhammad Ali anecdote and acknowledges financial support from Science  and  Engineering  Research Board (SERB) via the startup grant (File Number: YSS/2015/001696).
D.S.B acknowledges PhD fellowship support from  University Grants Commission (UGC) India. S.D. is grateful to the Indian National Science Academy and its Senior Scientist scheme for support and to IISER, Bhopal for its kind hospitality.
\bibliography{ref}
\appendix
\onecolumngrid
\section{Probability calculation}\label{appendixa}
For a telegraph noise we have (with $\eta(t)=\int_{0}^{t}\mathcal{F}(t^\prime) dt^\prime$)
\begin{equation}\label{eqa1}
i\eta(t)= -t(\gamma-i\epsilon).\mathcal{I} + t\sigma_z(\gamma\delta p + i\mu) + \gamma t(\sigma_x +i\delta p\sigma_y),
\end{equation}
where $\gamma = \frac{\lambda }{2}$ and $\delta p=(p_+-p_-)$.
Using a Pauli spin identity: $e^{i(\textbf{a}.\boldsymbol{\sigma})}=\mathcal{I}\cos{|\textbf{a}|} + i(\hat{n}.\vec{\sigma})\sin{|\textbf{a}|}$,
the exponential of Eqn.~\ref{eqa1} can be written as
\begin{equation}
e^{i\eta(t)}=\frac{1}{2} e^{-t(\gamma-i\epsilon)}\left[e^{\nu t}(1+\hat{h}.\vec{\sigma} + e^{-\nu t}(1+\hat{h}.\vec{\sigma})\right].
\end{equation}
Also, the conjugate equation is ($h_{x}^{\prime}=\gamma, h_{y}^\prime=-i\gamma\delta p, h_{z}^{\prime}=\gamma\delta p-i\mu$)
\begin{equation}
e^{-i\eta(t)}=\frac{1}{2} e^{-t(\gamma+i\epsilon)}\left[e^{\nu t}(1+\hat{h}^\prime.\vec{\sigma} + e^{-\nu t}(1+\hat{h}^\prime.\vec{\sigma})\right].
\end{equation}
Introducing $z=e^{ik}$ and $z^\prime = e^{ik^\prime}$, the expression for $(V_{k}^+(t)-V_{k^\prime}^+(t))$ can be solved to
\begin{eqnarray}
V_{k}^+(t)-V_{k^\prime}^+(t)= -\frac{\Delta}{8} e^{-\gamma t}\left\{(z-z^{\prime})\ e^{i\epsilon t}\left[e^{\nu t}(1+\hat{h}.\vec{\sigma})+e^{-\nu t}(1-\hat{h}.\vec{\sigma})\right]\right.\nonumber\\
\left. {} +(z^{*}-z^{\prime *})\ e^{-i\epsilon t}\left[e^{\nu t}(1+\hat{h}^\prime.\vec{\sigma})+e^{-\nu t}(1-\hat{h}^\prime.\vec{\sigma})\right]
\right\}.
\end{eqnarray}
Finally, we need to solve the integration 
\begin{eqnarray}
-i\int_{0}^{t} dt^\prime\left[V_{k}^+(t^\prime)-V_{k^\prime}^+(t^\prime)\right]= \frac{i\Delta}{8}\left\{(z-z^{\prime})\left[\frac{1-e^{-(\gamma-\nu)t+i\epsilon t}}{(\gamma-\nu)-i\epsilon}+\frac{1-e^{-(\gamma+\nu)t+i\epsilon t}}{(\gamma+\nu)-i\epsilon}\right]\qquad\qquad\quad\right.\nonumber\\
\left. {} +(z-z^{\prime})(\hat{h}.\vec{\sigma})\left[\frac{1-e^{-(\gamma-\nu)t+i\epsilon t}}{(\gamma-\nu)-i\epsilon}-\frac{1-e^{-(\gamma+\nu)t+i\epsilon t}}{(\gamma+\nu)-i\epsilon}\right]+c.c
\right\}.
\end{eqnarray}
Using the relations
\begin{equation}
\hat{h}.\vec{\sigma}=\frac{\gamma}{\nu}\sigma_x + \frac{i\gamma\delta p}{\nu}\sigma_y + \frac{(\gamma\delta p +i\mu)}{\nu}\sigma_z, \ \hat{h^\prime}.\vec{\sigma}=\frac{\gamma}{\nu^*}\sigma_x + \frac{i\gamma\delta p}{\nu^*}\sigma_y + \frac{(\gamma\delta p -i\mu)}{\nu^*}\sigma_z
\end{equation}
the exponential of the above equation can be written as
\begin{equation}
e^{-i\int_{0}^{t} dt^{\prime}(V_{k}^{+}(t^{\prime})-V_{k^{\prime}}^{+}(t^{\prime}))}=i\left[g_0(t).\mathcal{I} + g_1(t)\sigma_x + g_2(t)\sigma_y + g_3(t)\sigma_z\right],
\end{equation}
where
\begin{gather}
g_{0}(t)= \frac{\Delta}{8}\left\{(z-z^{\prime})\left[\frac{1-e^{-(\gamma-i\epsilon)t+\nu t}}{(\gamma-i\epsilon)-\nu}+\frac{1-e^{-(\gamma-i\epsilon)t-\nu t}}{(\gamma-i\epsilon)+\nu}\right]+(z^*-z^{\prime *})\left[\frac{1-e^{-(\gamma+i\epsilon)t+\nu^* t}}{(\gamma+i\epsilon)-\nu^*}+\frac{1-e^{-(\gamma+i\epsilon)t-\nu^* t}}{(\gamma+i\epsilon)+\nu^*}\right]
\right\}\quad\nonumber\\
g_{1}(t)= \frac{\Delta\gamma}{8}\left\{\frac{(z-z^{\prime})}{\nu}\left[\frac{1-e^{-(\gamma-i\epsilon)t+\nu t}}{(\gamma-i\epsilon)-\nu}-\frac{1-e^{-(\gamma-i\epsilon)t-\nu t}}{(\gamma-i\epsilon)+\nu}\right] +\frac{(z^*-z^{\prime *})}{\nu^*}\left[\frac{1-e^{-(\gamma+i\epsilon)t+\nu^* t}}{(\gamma+i\epsilon)-\nu^*}-\frac{1-e^{-(\gamma+i\epsilon)t-\nu^* t}}{(\gamma+i\epsilon)+\nu^*}\right]\right\}\quad\nonumber\\
g_{2}(t)= \frac{i\Delta\gamma\delta p}{8}\left\{\frac{(z-z^{\prime})}{\nu}\left[\frac{1-e^{-(\gamma-i\epsilon)t+\nu t}}{(\gamma-i\epsilon)-\nu}-\frac{1-e^{-(\gamma-i\epsilon)t-\nu t}}{(\gamma-i\epsilon)+\nu}\right] +\frac{(z^*-z^{\prime *})}{\nu^*}\left[\frac{1-e^{-(\gamma+i\epsilon)t+\nu^* t}}{(\gamma+i\epsilon)-\nu^*}-\frac{1-e^{-(\gamma+i\epsilon)t-\nu^* t}}{(\gamma+i\epsilon)+\nu^*}\right]
\right\}\nonumber\\
g_{3}(t)= \frac{\Delta\gamma\delta p}{8}\left\{\frac{(z-z^{\prime})}{\nu}\left[\frac{1-e^{-(\gamma-i\epsilon)t+\nu t}}{(\gamma-i\epsilon)-\nu}-\frac{1-e^{-(\gamma-i\epsilon)t-\nu t}}{(\gamma-i\epsilon)+\nu}\right] +\frac{(z^*-z^{\prime *})}{\nu^*}\left[\frac{1-e^{-(\gamma+i\epsilon)t+\nu^* t}}{(\gamma+i\epsilon)-\nu^*}-\frac{1-e^{-(\gamma+i\epsilon)t-\nu^* t}}{(\gamma+i\epsilon)+\nu^*}\right]
\right\} \nonumber\\ \quad + \frac{i\Delta\mu}{8}\left\{\frac{(z-z^{\prime})}{\nu}\left[\frac{1-e^{-(\gamma-i\epsilon)t+\nu t}}{(\gamma-i\epsilon)-\nu}-\frac{1-e^{-(\gamma-i\epsilon)t-\nu t}}{(\gamma-i\epsilon)+\nu}\right] -\frac{(z^*-z^{\prime *})}{\nu^*}\left[\frac{1-e^{-(\gamma+i\epsilon)t+\nu^* t}}{(\gamma+i\epsilon)-\nu^*}-\frac{1-e^{-(\gamma+i\epsilon)t-\nu^* t}}{(\gamma+i\epsilon)+\nu^*}\right]
\right\}\nonumber\\
\end{gather}
Also the expressions for $g_2(t)$ and $g_3(t)$ can be related to $g_1(t)=\alpha(t)$ as 
\begin{equation}
g_2(t) = i\delta p \alpha(t),\quad g_3(t) = \delta p \alpha(t) + \beta(t),
\end{equation}
where $\beta(t)$ is the second part of $g_3(t)$. The expression for $|\textbf{H}|$ can be solved to
\begin{equation}
|\textbf{H}| = \sqrt{\alpha^2(t) + \beta^2(t)+2\delta p \alpha(t)\beta(t)}.
\end{equation} 
Finally, substituting these into the expression for the probability propagator and taking the restricted averages~\citep{blume1968stochastic}, a simplified expression for the probability propagator for the case of telegraph noise can be obtained.  
\end{document}